\documentclass[final,numberedheadings]{aipproc}

\usepackage{amsmath}
\usepackage{amsfonts}
\usepackage{amssymb}

\layoutstyle{8x11single}

\begin{document}

\title{A brief introduction to cosmic topology}

\author{M.J. Rebou\c{c}as}{
  address={Centro Brasileiro de Pesquisas F\'{\i}sicas \\ 
           Rua Dr. Xavier Sigaud 150 \\
           22290-180 Rio de Janeiro - RJ, Brazil}  
                          }
\begin{abstract}
Whether we live in a spatially finite universe, and what its shape and 
size may be, are among the fundamental long-standing questions in
cosmology. These questions of topological nature have become particularly 
topical, given the wealth of increasingly accurate astro-cosmological 
observations, especially the recent observations of the cosmic microwave 
background radiation. An overview of the basic context of cosmic 
topology, the detectability constraints from recent observations, as well 
as the main methods for its detection and some recent results are 
presented.   
\end{abstract}

\maketitle

\section{Introduction}

Whether the universe is spatially finite and what is its shape 
and size are among the fundamental open problems that 
the modern cosmology seeks to resolve. These questions of 
topological nature have become particularly topical, given the 
wealth of increasingly accurate astro-cosmological observations, 
especially the recent observations of the cosmic microwave 
background radiation (CMBR)~\cite{WMAP}. 
An important point in these topological questions is that as 
a (local) metrical theory general relativity leaves 
the (global) topology of the universe undetermined. Despite this 
inability to predict the topology of the universe at a classical 
level, we should be able to devise strategies and methods to 
detect it by using data from current or future cosmological 
observations.

The aim of these lecture notes is to give a brief review of the 
main topics on cosmic topology addressed in four lectures
in the $\mathrm{XI}^{\mathrm{th}}$ Brazilian School of Cosmology 
and Gravitation, held in Mangaratiba, Rio de Janeiro from July 26 
to August 4, 2004.
Although the topics had been addressed with some details in 
the lectures, here we only intend to present a brief overview of 
the lectures. For more details we refer the readers to the long
list of references at the end of this article.

The outline of this article is a follows. In section~\ref{BasicContext}
we discuss how cosmic topology arises in the context of the 
standard Friedmann--Lema\^{\i}tre--Robertson--Walker (FLRW) cosmology, 
and what is the main observational physical effect used in the
search for a nontrivial topology of the spatial section of the 
universe. We also recall in this section some relevant
results about spherical and hyperbolic $3$-manifolds, which will be
useful in the following sections.
In section~\ref{Detect} we discuss the detectability of cosmic 
topology, present examples on how one can decide whether a 
given topology is detectable or not according to recent 
observations, and review some important results on this topic.
In section~\ref{StatMethods} we review the two main statistical 
methods to detect cosmic topology from the distribution of discrete 
cosmic sources.
In section~\ref{CMBmethods} we describe two methods devised for
the search of signs of a non-trivial topology in the CMBR maps.
 
\section{Basic Context}
\label{BasicContext}

General relativity (GR) relates the matter content of the
universe to its geometry, and reciprocally the geometry constrains 
the dynamics of the matter content. As GR is a purely metrical
(local) theory it does not fix the (global) topology of 
spacetime. 
To illustrate this point in a very simple way,
imagine a two-dimensional ($2$--D) world and its beings.
Suppose further these $2$--D creatures have a geometrical theory
of gravitation [an ($1+2$) spacetime theory], and modelling their
universe in the framework of this theory they found that 
the $2$--D geometry of the regular space is Euclidean --- they 
live in a spatially flat universe.
This knowledge, however, does not give them enough information
to determine the space topology of their world. Indeed, besides 
the simply-connected Euclidean plane $\mathbb{R}^2$, the space section of 
their universe can take either of the following multiply-connected 
space forms: the cylinder $\mathbb{C}^2 = \mathbb{R} \times \mathbb{S}^1$,
the torus $\mathbb{T}^2= \mathbb{S}^1 \times \mathbb{S}^1$, the  Klein
bottle $\mathbb{K}^2 = \mathbb{S}^1 \times \mathbb{S}^1_{\pi}$ and the 
M\"obius band $\mathbb{M}^2 = \mathbb{R} \times \mathbb{S}^1_{\pi}$. 
In brief, the local geometry constrains, but does not dictate the 
topology of the space. This is the very first 
origin of cosmic topology in the context of GR, as we shall discuss
in what follows.

Within the framework of the standard FLRW cosmology in the context
of GR,  the universe is modelled by a $4$~--~manifold $\mathcal{M}$ 
which is decomposed into $\mathcal{M} = R \times M$, and is endowed 
with a locally isotropic and homogeneous Robertson--Walker (RW) 
metric
\begin{equation}
\label{RWmetric} ds^2 = -dt^2 + a^2 (t) \left [ d \chi^2 +
f^2(\chi) (d\theta^2 + \sin^2 \theta  d\phi^2) \right ] \;,
\end{equation}
where $f(\chi)=(\chi\,$, $\sin\chi$, or $\sinh\chi)$ depending on
the sign of the constant spatial curvature ($k=0,1,-1$), and $a(t)$
is the scale factor.

\begin{figure}[!tbh]
\includegraphics[height=.2\textheight,width=0.3\textwidth]{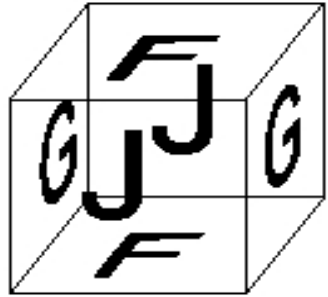}
\caption{A fundamental polyhedron of the Euclidean $3$--torus. The opposite 
faces are identified by the matching through translations of the pairs of 
equal opposite letters.}
\label{FigT3}
\end{figure}

The spatial section $M$ is usually taken to be one of the following
simply-connected spaces: Euclidean $\mathbb{E}^{3}$ ($k=0$), spherical 
$\mathbb{S}^{3}$ ($k=1$), or hyperbolic $\mathbb{H}^{3}$ ($k=-1$) spaces. 
However, since geometry does not dictate topology, the $3$-space $M$ may 
equally well be any one of the possible quotient manifolds 
$M = \widetilde{M}/\Gamma$, where $\Gamma$ is a discrete and fixed 
point-free group of isometries of the covering space 
$\widetilde{M}=(\mathbb{E}^3, \mathbb{S}^3, \mathbb{H}^3)$.
In forming the quotient manifolds $M$ the essential point is that they
are obtained from $\widetilde{M}$ by identifying points which are
equivalent under the action of the discrete group $\Gamma$. Hence,
each point on the quotient manifold $M$ represents all the
equivalent points on the covering manifold $\widetilde{M}$. The
action of $\Gamma$ tessellates (tiles) $\widetilde{M}$ into
identical cells or domains which are copies of what is known as
fundamental polyhedron (FP).
An example of quotient manifold in three dimensions is the $3$--torus 
$T^3=\mathbb{S}^1 \times \mathbb{S}^1 \times \mathbb{S}^1=\mathbb{E}^3/\Gamma$,
The  covering space clearly is $\mathbb{E}^3$, and a FP is a cube with 
opposite faces identified as indicated, in figure~\ref{FigT3}, by the equal 
opposite letters. This FP tiles the covering space $\mathbb{E}^3$. 
The group $\Gamma=\mathbb{Z} \times \mathbb{Z} \times \mathbb{Z}$ consists 
of discrete translations associated with the face identifications.

In a multiply connected manifold, any two given points may be joined
by more than one geodesic. Since the radiation emitted by cosmic
sources follows (space-time) geodesics, the immediate observational
consequence of a non-trivial spatial topology of $M$ is that the sky
may (potentially) show multiple images of radiating sources: cosmic
objects or specific correlated spots of the CMBR. At large
cosmological scales, the existence of these multiple images (or
pattern repetitions) is a physical effect often used in the
search for a nontrivial $3$-space topology.%
\footnote{Clearly we are assuming here that the radiation 
(light) must have sufficient time to reach the observer from 
multiple directions, or put in another way, that the universe 
is sufficiently 'small' so that this repetitions can be observed. In 
this case the observable horizon $\chi_{hor}$  exceeds at least 
the smallest characteristic size of $M$. A more detailed 
discussion on this point will be given in section~\ref{Detect}.}
In this article, in line with the usage in the literature, by 
cosmic topology  we mean the topology of the space section $M$ 
of the space-time manifold $\mathcal{M}$.

A question that arises at this point is whether one can use the 
topological multiple images of the same celestial objects such 
as cluster of galaxies, for example, to determine a nontrivial cosmic 
topology (see, e.g., refs~\cite{SokolovShvartsman1974}~--~\cite{Gomero2003a})
In practice, however, the identification of multiple images is
a formidable observational task to carry out because it involves 
a number of problems, some of which are:
\begin{itemize}
\item
Two images of a given cosmic object at different distances correspond 
to different periods of its life, and so they are in different stages 
of their evolutions, rendering problematic their identification as
multiple images.
\item
Images are seen from different angles (directions), which makes it 
very hard to recognize them as identical due to morphological 
effects;
\item
High obscuration regions or some other object can mask or even hide
the images;
\end{itemize}

These difficulties make clear that a direct search for multiples images is
not very promising, at least with available present-day technology. 
On the other hand, they motivate new search strategies and methods to 
determine (or just detect) the cosmic topology from observations. 
Before discussing in section~\ref{StatMethods} the statistical methods,
which have been devised to search for a possible nontrivial topology
from the distribution of discrete cosmic sources, we shall discuss in
the next section the condition for detectability of cosmic topology.

\section{Detectability of Cosmic Topology}
\label{Detect}

In this section we shall examine the detectability of cosmic 
topology problem for the nearly flat ($\Omega_0 \sim 1$) universes 
favored by current observation~\cite{SDSSWMAP}, and show that
a number of important results may be derived from the very fact 
that the cosmic topology is detectable. 
Thus, the results present in this section are rather general 
and hold regardless of the cosmic topology detection method
one uses, as long as it relies on images or pattern 
repetitions.

The extent to which a nontrivial topology of may or may not be 
detected for the current bounds on the cosmological density 
parameters has  been examined in a few articles~\cite{TopDetect}~--%
~\cite{WeLeUz2003}. The discussion below is based upon our 
contribution to this issue~\cite{TopDetect}~--~\cite{MaMoRe2004}.  

In order to state the conditions for the detectability of cosmic
topology in the context of standard cosmology, we note that for
non-flat metrics of the form~(\ref{RWmetric}), the scale factor
$a(t)$ can be identified with the curvature radius of the spatial
section of the universe at time $t$. Therefore $\chi$ is the
distance of a point $p =(\chi, \theta, \phi)$ to the coordinate 
origin $O$ (in the covering space) in units of the curvature
radius, which is a natural unit of length that shall be used
throughout this paper.

\begin{figure}[tbh]
\includegraphics[height=.3\textheight]{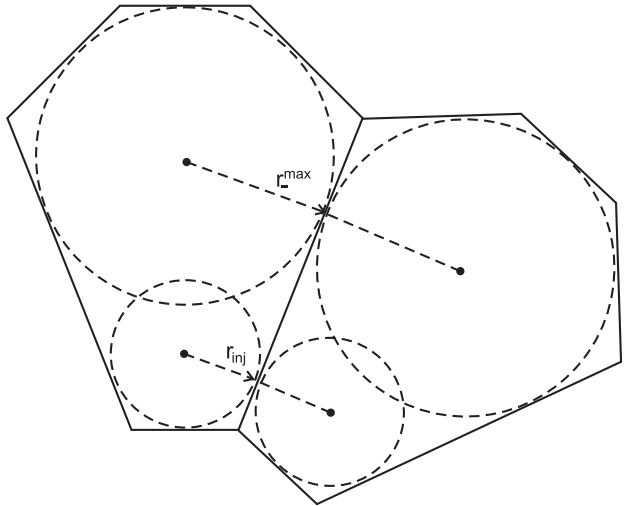}
\caption{A schematic representation of two fundamental cells,
and the indication of the injectivity radius $r_{inj}$, which
is the radius of the smallest sphere `inscribable' in the 
fundamental domain. The  radius of the largest sphere 
`inscribable' $r^{max}$ is also shown.}
\label{fig:r_inj}
\end{figure}

The study of the detectability of a possible non-trivial topology of
the spatial section $M$ requires a topological typical length which
can be put into correspondence with observation survey depths
$\chi_{obs}$ up to a redshift $z=z_{obs}$. A suitable characteristic
size of $M$, which we shall use in this paper, is the so-called
injectivity radius $r_{inj}$, which is nothing but the radius of the
smallest sphere `inscribable' in $M$, and is defined in terms of the
length of the smallest closed geodesics $\ell_M\,$ by $r_{inj} =
\ell_M/2$ (see fig.~\ref{fig:r_inj}).

Now, for a given survey depth $\chi_{obs}$ a topology is said to be
undetectable if $\chi_{obs} < r_{inj}$. In this case no multiple 
images (or pattern repetitions of CMBR spots) can be detected in 
the survey of depth $\chi_{obs}$. On the other hand, when 
$\chi_{obs} > r_{inj}$, then the topology is detectable in 
principle or potentially detectable.

In a globally homogeneous manifold the above detectability condition
holds regardless of the observer's position, and so if the topology
is potentially detectable (or is undetectable) by an observer at 
$x \in M$, it is potentially detectable (or is undetectable) by an
observer at any other point in the $3$-space $M$. 
However, in globally inhomogeneous manifolds the detectability of 
cosmic topology depends on both the observer's position $x$ and 
the survey depth $\chi_{obs}$. Nevertheless, even for globally 
inhomogeneous manifolds the above defined `global' injectivity 
radius $r_{inj}$ can be used to state an \emph{absolute undetectability} 
condition, namely $r_{inj} > \chi_{obs}$, in the sense that if
this condition holds the topology is undetectable for any
observer at any point in $M$.
Reciprocally, the condition $\chi_{obs} > r_{inj}$ allows potential 
detectability (or detectability in principle) in the sense that, if 
this condition holds, multiple images of topological origin are 
potentially observable at least for some observers suitably located 
in $M$. 
An important point is that for spherical and hyperbolic manifolds,
the `global' injectivity radius $r_{inj}$ expressed in terms of the 
curvature radius, is a constant (topological invariant) for a given 
topology.

Before proceeding further we shall recall some relevant results 
about spherical and hyperbolic $3$--manifolds, which will be
used to illustrate the above detectability condition. 
The multiply connected spherical $3$-manifolds are of the form 
$M=\mathbb{S}^3/\Gamma$, where $\Gamma$ is a finite subgroup of $SO(4)$ 
acting freely on the $3$-sphere. These manifolds were originally 
classified  by Threlfall and Seifert~\cite{ThrelfallSeifert}, 
and are also discussed by Wolf~\cite{Wolf} (for a description 
in the context of cosmic topology see~\cite{Ellis71}). 
Such a classification consists essentially in the enumeration of 
all finite groups $\Gamma \subset SO(4)$, and then in grouping 
the possible manifolds in classes. 
In a recent paper, Gausmann \emph{et al.\/}~\cite{GLLUW} recast the
classification in terms of single action, double action, and linked
action manifolds. In table~\ref{SingleAction} we list the single
action manifolds together with the symbol often used to refer to
them, as well as the order of the covering group $\Gamma$ and the
corresponding injectivity radius. It is known that single action
manifolds are globally homogeneous, and thus the detectability
conditions for an observer at an arbitrary point $p \in M$ also hold
for an observer at any other point $q \in M$. Finally we note that
the binary icosahedral group $I^{\ast}$ gives the known Poincar\'e
dodecahedral space, whose fundamental polyhedron is a regular
spherical dodecahedron, $120$ of which tile the $3$-sphere into
identical cells which are copies of the FP.

\begin{table}[!htb]
\begin{tabular}{lcc}
\hline
\tablehead{1}{l}{b}{\ Name \&  Symbol \ }
&\tablehead{1}{c}{b}{\ Order of $\,\Gamma$ \ }
&\tablehead{1}{c}{b}{\ Injectivity Radius \ } \\
\hline 
Cyclic              $\;Z_n$   & $n$  & $\pi/n$           \\
Binary dihedral     $\;D^*_m$ & $4m$ & $\pi / 2m $      \\
Binary tetrahedral  $\;T^*$   & 24   & $ \pi/6$          \\
Binary octahedral   $\;O^*$   & 48   & $\pi/8$          \\
Binary icosahedral  $\;I^*$   & 120  & $\pi/10$         \\
\hline
\end{tabular}
\caption{Single action spherical manifolds together with the order
of the covering group and the injectivity radius.}
\label{SingleAction}
\end{table}

Despite the enormous advances made in the last few decades, there is
at present no complete classification of hyperbolic $3$-manifolds.
However, a number of important results have been obtained, including
the two important theorems of Mostow~\cite{Mostow} and
Thurston~\cite{Thurston}. According to the former, geometrical
quantities of orientable hyperbolic manifolds, such as their volumes
and the lengths of their closed geodesics, are topological
invariants. Therefore quantities such as the `global' injectivity
radius $r_{inj}$ (expressed in units of the curvature radius) are
fixed for each manifold. Clearly this property also holds for
spherical manifolds. 

According to Thurston's theorem, there is a countable infinity of
sequences of compact orientable hyperbolic manifolds, with the
manifolds of each sequence being ordered in terms of their volumes.
Moreover, each sequence has as an accumulation point a cusped
manifold, which has finite volume, is non-compact, and has
infinitely long cusped corners~\cite{Thurston}.

\begin{table}[!htb]
\begin{tabular}{lcc}
\tablehead{1}{l}{b}{\ \ \ Manifold \ \ \ }
&\tablehead{1}{c}{b}{\ \ \ Injectivity Radius  } 
&\tablehead{1}{c}{b}{\ \ \ Volume \ \ \ } \\
\hline 
\ \ m003(-4,1)& 0.177  & 1.424 \\ 
\ \ m004(3,2) & 0.181  & 1.441  \\ 
\ \ m003(-3,4)& 0.182  & 1.415   \\ 
\ \ m004(1,2) & 0.183  & 1.398  \\ 
\ \ m004(6,1) & 0.240  & 1.284  \\ 
\ \ m003(-4,3)& 0.287  & 1.264  \\ 
\ \ m003(-2,3)& 0.289  & 0.981  \\ 
\ \ m003(-3,1)& 0.292  & 0.943   \\ 
\ \ m009(4,1) & 0.397  & 1.414  \\ 
\ \ m007(3,1) & 0.416  & 1.015  \\ \hline
\end{tabular}
\caption{First seven manifolds in the Hodgson-Weeks
census of closed hyperbolic manifolds, ordered by the
injectivity radius $r_{inj}$, together with their
corresponding volume.}
\label{10HW-Census}
\end{table}

Closed orientable hyperbolic $3$-manifolds can be constructed from
these cusped manifolds. The compact manifolds are obtained through a
so-called Dehn surgery which is a formal procedure identified by two
coprime integers, i.e. winding numbers $(n_1,n_2)$. These manifolds
can be constructed and
studied with the publicly available software package SnapPea%
~\cite{SnapPea}. SnapPea names manifolds according to the seed
cusped manifold and the winding numbers. So, for example, the
smallest volume hyperbolic manifold known to date (Weeks' manifold)
is named as m$003(-3,1)$, where m003 corresponds to a seed cusped
manifold, and $(-3,1)$ is a pair of winding numbers. Hodgson and
Weeks~\cite{SnapPea,HodgsonWeeks} have compiled a census containing
$11031$ orientable closed hyperbolic 3-manifolds ordered by
increasing volumes. In table~\ref{10HW-Census} we collect the first
ten manifolds from this census with the lowest volumes, ordered by
increasing injectivity radius $r_{inj}$, together with their
volumes.

To illustrate now the above condition for detectability
(undetectability) of cosmic topology, in the light of recent 
observations~\cite{SDSSWMAP} we assume 
that the matter content of the universe is well approximated 
by dust of density $\rho_m$ plus a cosmological constant 
$\Lambda$. In this cosmological setting the current curvature radius 
$a_0$ of the spatial section is related to the total density 
parameter $\Omega_0$ through the equation 
\begin{equation}
\label{CurvRad}
a_0^2 = \frac{k}{H_0^2(\Omega_0-1)} \; ,
\end{equation}
where $H_0$ is the Hubble constant, $k$ is the normalized 
spatial curvature of the RW metric~(\ref{RWmetric}), and
where here and in what follow the subscript $0$ denotes 
evaluation at present time $t_0$. Furthermore, in this
context the redshift-distance relation in units of the 
curvature radius, $a_0=R(t_0)$, reduces to 
\begin{equation}
\label{redshift-dist}
\chi(z) = \sqrt{|1-\Omega_0|} \int_1^{1+z} \hspace{-4mm}
\frac{dx}{\sqrt{x^3 \Omega_{m0} + x^2 (1- \Omega_0) + 
 \Omega_{\Lambda 0}}} \; ,
\end{equation}
where $\Omega_{m0}$ and $\Omega_{\Lambda 0}$ are, respectively,
the matter and the cosmological density parameters, and 
$\Omega_0 \equiv \Omega_{m0} + \Omega_{\Lambda 0}$.
For simplicity, on the left hand side of~(\ref{redshift-dist})
and in many places of this article, we have 
left implicit the dependence of the function $\chi$ on the 
density components.

\begin{figure}[!tbh]
\includegraphics[height=.35\textheight,width=0.4\textwidth]{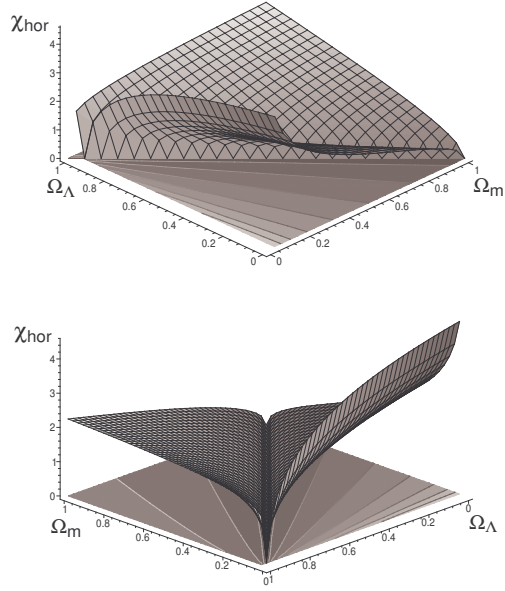}
\caption{The behaviour of $\chi_{hor}=\chi(\Omega_{m0},\Omega_{\Lambda0},z)$, 
in units of curvature radius, for $z=1100$ as a function of 
the density parameters $\Omega_{\Lambda 0}$ and $\Omega_{m0}\,$.
These f\/igures show clearly the rapid way $\chi_{hor}$ falls off to 
zero for nearly f\/lat (hyperbolic or elliptic) universes}
\label{fig:Bird}
\end{figure}

A first qualitative estimate of the constraints on detectability of 
cosmic topology from nearflatness can be obtained from the function
$\chi(\Omega_{m0},\Omega_{\Lambda0},z)\,$ given by~(\ref{redshift-dist})
for a fixed survey depth $z$. Figure~\ref{fig:Bird} clearly demonstrates 
the rapid way $\chi$ drops to zero in a narrow neighbourhood of the 
$\Omega_0 = 1$ line. This can be understood intuitively from~(\ref{CurvRad}), 
since the natural unit of length (the curvature radius $a_0$) goes 
to infinity as $\Omega_0 \to 1$, and therefore the depth $\chi$ (for 
any fixed $z$) of the observable universe becomes smaller in this 
limit.
{}From the observational point of view, this shows that the detection of 
the topology of the nearly flat universes becomes more and more difficult 
as $\Omega_0 \to 1$, a limiting value favoured by recent observations.
As a consequence, by using any method which relies on observations of 
repeated patterns the topology of an increasing number of nearly flat 
universes becomes undetectable in the light of the recent observations,
which indicate that $\Omega_0 \sim 1$.  

{}From the above discussion it is clear that cosmic topology 
may be undetectable for a given survey up to a depth $z_{max}$, 
but detectable if one uses a deeper survey. At present the deepest 
survey available corresponds to $z_{max}=z_{LSS} \sim 1\,000$,
with associated depth $\chi(z_{LSS})$. So the most promising 
searches for cosmic topology through multiple images of radiating 
sources are based on CMBR.

As a concrete quantitatively example we consider universes with
that possess a cyclic single action spherical topologies 
for $\Omega_0 = 1.08$ and $\Omega_{\Lambda} = 0.66$. From table%
~\ref{SingleAction} we have  $r_{inj} = \pi/n$ which together with
the undetectability condition give
\begin{equation}
\chi_{obs} < \,r_{inj} \quad \Longrightarrow \quad 
n < n^{*}= \,\mbox{int}\,\,( \frac{\pi}{\chi_{obs}})\;,
\end{equation}
where $\mbox{int}\,(x)$ denotes the integer part of $x$.

In table~\ref{Tb:UndSingleAct} for distinct redshifts $z_{max}$ we
collect the corresponding survey depth $\chi_{obs}$ and the limiting
value below which the cyclic single action manifold is undetectable.
According to this table the cyclic group manifolds $\mathbb{Z}_2$ and 
$\mathbb{Z}_3$ are undetectable even if one uses CMBR, while the 
manifolds $S^3/\mathbb{Z}_p$ for $p \geq 4$ are detectable with CMBR. 
For the same values of the density parameters (besides $\mathbb{Z}_2$ 
and $\mathbb{Z}_3$ manifolds) the manifolds $\mathbb{Z}_4$, $\mathbb{Z}_5$ 
and $\mathbb{Z}_6$ are undetectable using sources of redshifts up to 
$z_{max}=6$.

\begin{table}[!htb]
\begin{tabular}{ccc}
\tablehead{1}{c}{b}{\ \ \ Redshift $\,z_{max}$ \ \ \ }
&\tablehead{1}{c}{b}{\ \ \ Depth $\,\chi_{obs}$ \ \ \ } 
&\tablehead{1}{c}{b}{\ \ \ Limiting value $\,n^*$ \ \ \ } \\
\hline 
  1100   & 0.811 & 4 \\ 
    6    & 0.496 & 7 \\ 
    1    & 0.209 & 16 \\ \hline
\end{tabular}
\caption{For each $z_{max}$ the corresponding values
$\chi_{obs}$ for $\Omega_0 = 1.08$ and $\Omega_{\Lambda} = 0.66$. The
integer number $n^*$ is the limiting value below which the corresponding
cyclic topology is undetectable.}
\label{Tb:UndSingleAct}
\end{table}

To quantitatively illustrate the above features of the detectability
problem in the hyperbolic case ($\Omega_0< 1$) , we shall examine the 
detectability of cosmic topology of the first ten smallest (volume) 
hyperbolic universes.
To this end we shall take the following interval of the density 
parameters values consistent with current observations:
$\Omega_0 \in [0.99,1)$ and $\Omega_{\Lambda 0} \in [0.63,0.73]$.
In this hyperbolic sub-interval one can calculate the largest 
value of $\chi_{obs}(\Omega_{m0},\Omega_{\Lambda0},z)\,$ for the 
last scattering surface ($z=1\,100$), and compare with the injectivity 
radii $r_{inj}$ to decide upon detectability. From~(\ref{redshift-dist})
one obtains $\chi^{max}_{obs} = 0.337\,$.

Table~\ref{Tb:10smallest} summarizes the results for CMBR ($z=1\,100$),
which have been refined upon by Weeks~\cite{Weeks2003}. 
It makes explicit the very important fact that there are undetectable 
topologies by any methods that rely on pattern repetitions even if one 
uses CMBR, which corresponds to the deepest survey depth $\chi(z_{LSS})$.

\begin{table}[!htb]
\begin{tabular}{lccc}
\tablehead{1}{l}{b}{\ \ \ Manifold \ \ \ }
&\tablehead{1}{c}{b}{\ \ \ Volume \ \ \  }
&\tablehead{1}{c}{b}{Injectivity radius} 
&\tablehead{1}{c}{b}{\ \ \ Detectability with CMBR \ \ \ } \\
\hline 
\ \ m003(-3,1) &  0.943  & 0.292  & Potential Detectable \\ 
\ \ m003(-2,3) &  0.981  & 0.289  & Potential Detectable \\ 
\ \ m007(3,1)  &  1.015  & 0.416  & Undetectable \\
\ \ m003(-4,3) &  1.264  & 0.287  & Potential Detectable \\ 
\ \ m004(6,1)  &  1.284  & 0.240  & Potential Detectable \\ 
\ \ m004(1,2)  &  1.398  & 0.183  & Potential Detectable \\ 
\ \ m009(4,1)  &  1.414  & 0.397  & Undetectable \\
\ \ m003(-3,4) &  1.415  & 0.182  & Potential Detectable \\ 
\ \ m003(-4,1) &  1.424  & 0.177  & Potential Detectable \\ 
\ \ m004(3,2)  &  1.441  & 0.181  & Potential Detectable \\ \hline 
\end{tabular}
\caption{Restrictions on detectability of cosmic topology
for $\Omega_0\!\!=\!0.99 \;\mbox{with}\;
\Omega_{\Lambda 0} \in [0.63,0.73] $
for the first ten smallest known hyperbolic manifolds.
A survey depth corresponding to CBMR ($z_{max}=1100$) was
used. The manifolds are ordered by increasing volumes.}
\label{Tb:10smallest}
\end{table}

Hitherto we have considered the detectability of nearly flat
universe, but one can alternatively ask what is the region
of the density parameter spaces for which topologies are
detectable. To this end, we note that for a given (fixed) survey 
with redshift cut-off $z_{obs}$, and for a given manifold
with injectivity radius $r_{inj}^{\ M}$ one can solve the 
equation
\begin{equation} \label{XobsRinj}
\chi (\Omega_0, \Omega_{\Lambda},z_{obs}\,) = \, r_{inj}^{\ M} \;,
\end{equation}
which amounts to finding pairs ($\Omega_0,\Omega_{\Lambda0}$)
in the density parameter $\Omega_0\,$--$\,\,\Omega_{\Lambda 0}$
plane for which eq.~(\ref{XobsRinj}) holds.%
\footnote{Since $\Omega_0 = \Omega_{m0}+\Omega_{\Lambda0}$ 
we can clearly take $\chi$ as function of either ($\Omega_0,\Omega_{m0}$) 
or ($\Omega_{m0},\Omega_{\Lambda0}$).}  

Consider now the set of the 19 smallest manifolds of the 
Hodgson-Weeks census in conjunction with the hyperbolic 
region
\begin{equation}
\label{hyper-data2}
\Omega_0 \in [0.98,1) \qquad \mbox{and} \qquad
\Omega_{\Lambda 0} \in [0.62,0.79] \;,
\end{equation}
and the eqs.~(\ref{redshift-dist}) and~(\ref{XobsRinj}). 
The manifold in this set with the lowest $r_{inj} (= 0.152)$ 
is $m003(-5,4)\,$ (see~\cite{SnapPea}).
Figure~\ref{fig:UndetecReg} gives the solution curve of 
equation~(\ref{XobsRinj}) in the 
$\Omega_0\,$--$\,\,\Omega_{\Lambda 0}$ plane for 
$r_{inj}=0.152$ and $r_{inj}= 0.081$,
where a survey of depth $z_{max}=1\,100$ (CMBR) was used.
This figure also contains a dashed rectangular box, 
representing the relevant part 
of the recent hyperbolic region~({\ref{hyper-data2}).
\begin{figure}[!htb]
\includegraphics[height=.3\textheight]{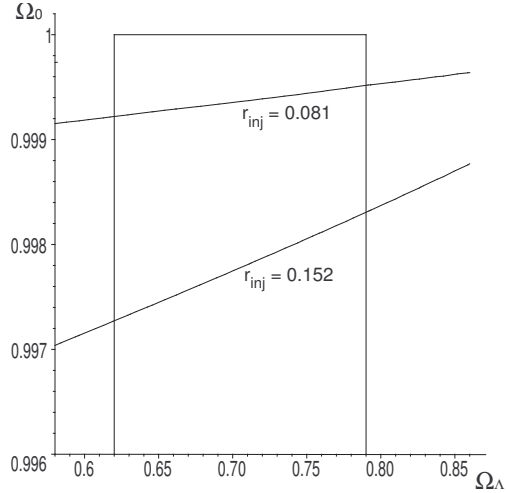}
\caption{The solution curves of $\chi_{obs}(\Omega_0,\Omega_{\Lambda 0},z) 
= r_{inj}$, as plots of $\Omega_0$ versus $\Omega_{\Lambda 0}$  for
$r_{inj}=0.081$ and $r_{inj} = 0.152$ . A survey with depth 
$z_{max}= 1100$ (CMBR) was used. 
The dashed rectangular box represents the relevant part 
of the hyperbolic region~(\ref{hyper-data2}) of the 
parameter space given by recent observations.   
The undetectable regions of the parameter space 
($\Omega_0, \Omega_{\Lambda 0}$), corresponding to each 
value of $ r_{inj}$,  lie above the related curve.}
\label{fig:UndetecReg}
\end{figure}
For each value of $r_{inj}$ undetectability is ensured 
for the values of cosmological parameters (region in 
the $\Omega_{0}\,$--$\,\,\Omega_{\Lambda 0}$ plane) 
which lie above the corresponding solution curve 
of~(\ref{XobsRinj}).
Thus, considering the solution curve of~(\ref{XobsRinj}) 
for $r_{inj} = 0.152$, for example, one finds that the topology 
of none of the 19 smallest manifolds of the census would be 
detectable, if $\Omega_0 \gtrsim 0.9971$. On the other hand,
one has that this value is a lower bound for the total density
parameter $\Omega_0$ if it turns out that one of these 19
hyperbolic manifolds is detected. 

For a given survey with redshift cut-off $z_{obs}$, the redshift
distance function $\chi_{obs}$ clearly depends on the way one
models the matter-energy content of the universe. This point
has been recently discussed by Mota \emph{et al.} in a unifying 
dark matter and dark energy framework~\cite{MaMoRe2004}.

In what follows  we shall briefly discuss two 
important results related to detectability of cosmic topology.
For more details we refer the readers to refs.~\cite{MotaRebTav03}
and~\cite{MotaRebTav04}.  
Regarding the first, consider again the solution curve 
of~(\ref{XobsRinj}) in the parameter plane. For a given survey 
depth $z_{obs}$ we define the secant line as the line joining 
the points $( \widetilde{\Omega}_{m0},0)$ and 
$(0,\widetilde{\Omega}_{\Lambda0})$ where the contour curve 
intersects the axes $\Omega_{m0}$ and $\Omega_{\Lambda0}$, 
respectively. Clearly the equation of this line is given 
by
\begin{equation}   \label{crit-extr}
\frac{\Omega_{m0}}{\widetilde{\Omega}_{m0}}+\frac{\Omega_{\Lambda0}}%
{\widetilde{\Omega}_{\Lambda0}}=1\;.
\end{equation}
It is possible to show that the solution curve of~(\ref{XobsRinj}) is
convex and concave, respectively, in the hyperbolic and spherical
regions of the parameter plane $\Omega_{\Lambda0}$~--~$\Omega_{m0}$.
This property can also be gleaned from the parametric plot of
the solution curve of~(\ref{XobsRinj}), and ensures that the secant 
line crosses the contour line only at the $\Omega_{m0}$ and 
$\Omega_{\Lambda0}$ axes. As a consequence the region between the
secant line and the flat line $\Omega_{\Lambda0}+\Omega_{m0}=1$
lies inside the undetectability region of the parameter plane.
Thus, the secant line approximation to the solution curve 
of~(\ref{XobsRinj}) gives a sufficient condition for undetectability
of the corresponding topology with injectivity radius $r_{inj}^{\ M}$.
A closed form for this sufficient condition can be obtained from
(\ref{crit-extr}) in the limiting case $z\rightarrow\infty$. 
As a result one has that a universe with space section M 
has undetectable topology if

\parbox{14cm}{\begin{eqnarray*}
\cosh^{2} \,(\, r_{inj}^{\:M} \,/\, 2 \,)\; \Omega_{m0} +
\Omega_{\Lambda0} & >& 1\;, \quad \mathrm{for}
\;\quad \Omega_{0}<1\;,  \nonumber\\
\cos^{2}\,(\, r_{inj}^{\:M} \,/\, 2 \,)\; \Omega_{m0}+
\Omega_{\Lambda0} & < &1\;, \quad \mathrm{for}
\;\quad \Omega_{0}>1\;.
                \end{eqnarray*}}  \hfill
\parbox{1cm}{\begin{eqnarray}   \label{analit}  \end{eqnarray}}
Despite its simple form, this result is of considerable interest in that
it gives a test for undetectability for \emph{any} $z$.

The condition~(\ref{analit}) can easily be written in terms of
either $\Omega_0$ and $\Omega_{\Lambda 0}$ or $\Omega_0$ and
$\Omega_{m0}$~\cite{MotaRebTav03}. So, for example, a universe
space section M has undetectable topology if

\parbox{14cm}{\begin{eqnarray*}
\Omega_{0} & > & 1 \,
-\,\sinh^{2}\,(\,r_{inj}^{M}\,/\,2\,)\,\; \Omega_{m0}\;, \quad
\mathrm{for} \quad \Omega_{0}<1\;, \nonumber \\ \Omega_{0} & < &
1 \: + \:\, \sin^2\,(\,r_{inj}^{M}\,/\,2\,)
\;\: \Omega_{\,m0}\;,\quad \: \mathrm{for}\quad \Omega_{0}>1 \;.
            \end{eqnarray*}}  \hfill
\parbox{1cm}{\begin{eqnarray}   \label{analit3} \end{eqnarray}}

{}From table~\ref{SingleAction} we have the injectivity radius
for the single action cyclic $\mathbb{Z}_n$ and binary dihedral $D_m^*$
families are given, respectively, by $r_{inj}=\pi/n$ and
$r_{inj}=\pi/2m$. This allows to solve the equation corresponding
to~(\ref{analit3}) to obtain

\parbox{14cm}{\begin{eqnarray*}
n^{\ast} & = &\mathrm{int}\left\{\,\frac{\pi}{2}\,\left[\,\arcsin\sqrt{\,
\frac{\Omega_{0}-1}
 {\Omega_{m0}}\;}\:\,\right]^{-1} \:\right\} \;, \nonumber \\
 m^{\ast} & = &\mathrm{int}\left\{\,\frac{\pi}{4}\,\left[\,\arcsin\sqrt{\,
\frac{\Omega_{0}-1}
 {\Omega_{m0}}\;}\:\,\right]^{-1}\:\right\} \;,
 \end{eqnarray*} }  \hfill
\parbox{1cm}{\begin{eqnarray} \label{m*} \end{eqnarray} }
where $\mathrm{int}[x]$ denotes the integer part of $x$.
Thus, for these two classes of manifold there is always
$n^*$ and $m^*$ such that the corresponding topology is
detectable for $n > n^*$ and $m > m^*$, given in terms
of the density parameters.

The second important result is related to detectability of very 
nearly flat universes, for which $|\Omega_{0}-1|\ll 1$~\cite{MotaRebTav04}.
If in addition to this condition we make two further physically 
motivated assumptions: (i) the observer is at a position $x$ 
where the topology is detectable, i.e. $r_{inj}(x)<\chi_{obs}$;
and (ii) the topology is not excludable, i.e. it does not produce 
too many images so as to be ruled out by present observations. 
Thus, these main physical assumption can be summarized as
\begin{equation}
r_{inj}(x)\,\lesssim\, \chi_{obs}\, \ll \, 1\;.\label{hip}%
\end{equation}
These assumptions severely restricts the set of detectable
nearly flat manifolds. Thus in the case of spherical manifolds, only
cyclic ($r_{inj}=\pi/n$) and binary dihedrical spaces ($r_{inj}=\pi/2m$) 
of sufficiently high order of $n$ or $4m$ are detectable. 
In the hyperbolic case, the only detectable manifolds are the so-called 
nearly cusped manifolds, which are sufficiently similar to the cusped 
manifolds (cusped manifolds are non-compact, and possess regions with 
arbitrarily small $r_{inj}(x)$.)

In a recent study~\cite{MotaRebTav04} we considered both classes of
manifolds and showed that a generic detectable spherical or
hyperbolic manifold is locally indistinguishable from either a
cylindrical ($\mathbb{R}^{2} \times\mathbb{S}^{1}$) or toroidal
($\mathbb{R}\times\mathbb{T}^{2}$) manifold, irrespective of its
global shape. 
These results have important consequences in the development of search
strategies for cosmic topology. They show that for a typical
observer in a very nearly flat universe, the 'detectable part' of
the topology would be indistinguishable from either
$\mathbb{R}^{2} \times\mathbb{S}^{1}$ or $\mathbb{R}
\times\mathbb{T}^{2}$ manifold. 

To conclude this section, we mention that Makler \emph{et al.} have
examined, in a recent article~\cite{MaMoRe2005}, the extent to what  
a possible detection of a non-trivial topology of a low curvature 
($\Omega_0 \sim 1$) universe may be used to place constraints on the
matter content of the universe, focusing our attention on the
generalized Chaplgygin gas (GCG) model, which unifies dark matter
and dark energy in a single matter component. It is shown that besides
constraining the GCG parameters, the detection of a nontrivial
topology also allows to set bounds on the total density parameter
$\Omega_0$. It is also studied the combination of the bounds from the
topology detection  with the limits that arise from  current data on SNIa, 
and  shown that the detection of a nontrivial topology sets
complementary bounds on the GCG parameters (and on $\Omega_0$) to
those obtained from the SNIa data alone (for examples of local physical 
effect of a possible nontrivial topology see, e.g., refs.%
~\cite{ORT1994}~--~\cite{MuFaOp2002}).

\section{Pair Separations Statistical methods} 
\label{StatMethods} 

On the one hand the most fundamental consequence of a multiply connected
spatial section $M$ for the universe is the existence of multiple 
images of cosmic sources, on the other hand a number of observational 
problems render the direct identification of these images practically 
impossible. 
In the statistical approaches to detect the cosmic topology instead of 
focusing on the direct recognition of multiple images, one treats 
statistically the images of a given cosmic source, and use (statistical) 
indicators or signatures in the search for a sign of a nontrivial 
topology. Hence the statistical methods are not plagued by direct 
recognition difficulties such as morphological effects, and distinct 
stages of the evolution of cosmic sources.
 
The key point of these methods is that in a universe with detectable 
nontrivial topology at least one of the characteristic sizes
of the space section $M$ is smaller than a given  
survey depth $\chi_{obs}$, so the sky should show multiple images of
sources, whose $3$--D positions are correlated by the isometries  
of the covering group $\Gamma$. These methods rely on the fact that 
the correlations among the positions of these images can be couched 
in terms of distance correlations between the images, and use 
statistical indicators to find out signs of a possible nontrivial 
topology of $M$. 

In 1996 Lehoucq \emph{et al.\/}~\cite{LeLaLu1996} proposed the 
first statistical method (often referred to as cosmic crystallography), 
which looks for these correlations by using pair separations histograms 
(PSH). To build a PSH we simply evaluate a suitable one-to-one function 
$F$ of the distance $d$ between a pair of images in a catalogue $\mathcal{C}$, 
and define $F(d)$ as the pair separation: $s = F(d)$. Then we depict 
the number of pairs whose separation lie within certain sub-intervals $J_i$ 
partitions of $( 0,\,s_{max} ]$, where $s_{max} = F(2\chi_{max})$, and 
$\chi_{max}$ is the survey depth of $\mathcal{C}$. A PSH is just a 
normalized plot of this counting. In most applications in the 
literature the separation is taken to be simply the distance between 
the pair $s=d$ or its square $s=d^2$, $J_i$ being, respectively, a 
partition of $(0, 2 \chi_{max}]$ and $(0, 4^{}_{} \chi_{max}^2]$. 

The PSH building procedure can be formalized as follows. Consider a
catalogue $\mathcal{C}$ with $n$ cosmic sources and denote by
$\eta(s)$ the number of pairs of sources whose separation is $s$.
Divide the interval $(0,s_{max}]$ in $m$ equal sub-intervals
(bins) of length    $ \delta s = s_{max} / {m}$, being        
\[
J_i = (s_i - \frac{\delta s}{2} \, , \, s_i + \frac{\delta
s}{2}] \;,; \quad  i=1,2, \dots ,m \;\,, \quad
\]
and centered at $ s_i = \,(i - \frac{1}{2})\, \delta s \,$.
The PSH is defined as the following counting function: 
\begin{equation}
\label{PSH}
\Phi(s_i)=\frac{2}{n(n-1)}\,\,\frac{1}{\delta s}\,
               \sum_{s \in J_i} \eta(s) \; ,
\end{equation}
which can be seen to be subject to the normalization condition
$\sum_{i=1}^m \Phi(s_i)\,\, \delta s = 1 \;.$
An important advantage of using \emph{normalized\/} PSH's is that
one can compare histograms built up from catalogues with
dif\/ferent number of sources. 

An example of PSH obtained through simulation for a universe 
with nontrivial topology is given in Fig.~\ref{fig:PSH-T3}. 
Two important features should be noticed: (i) the presence of 
the very sharp peaks (called spikes); and (ii) the existence of a
'mean curve' above which the spikes stands. This curve 
corresponds to an expected pair separation histogram (EPSH) 
$\Phi_{exp}(s_i)$, which is a typical PSH from which the 
statistical noise has been withdrawn, that is 
$\Phi_{exp}(s_i) = \Phi(s_i)  - \rho(s_i)\,$, where $\rho(s_i)$ 
represents the statistical f\/luctuation that arises in the 
PSH $\Phi(s_i)$.  
 
\begin{figure}[!htb] 
\includegraphics[height=.3\textheight]{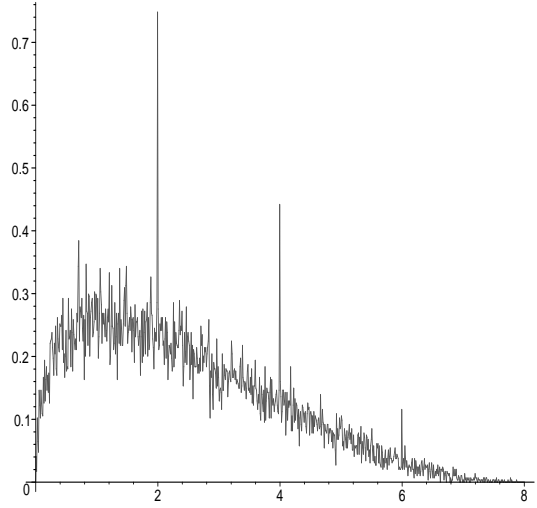}
\caption{Typical PSH for a flat universe with
a $3$--torus topology. The horizontal axis gives the squared pair
separation $s^2$, while the vertical axis provides a normalized
number of pairs.} 
\label{fig:PSH-T3}
\end{figure} 

The primary expectation was that the distance correlations 
would manifest as topological spikes in PSH's, and that 
the spike spectrum of topological origin would be a definite 
signature of the topology~\cite{LeLaLu1996}. 
While the first simulations carried out for specific flat manifolds 
appeared to confirm this expectation~\cite{LeLaLu1996}, 
histograms subsequently generated for specific hyperbolic 
manifolds revealed that the corresponding PSH's exhibit no spikes%
~\cite{LeLuUz1999,FagGaus1998}.
Concomitantly, a theoretical statistical analysis of the distance 
correlations in PSH's was accomplished, and a proof was presented 
that the spikes of topological origin in PSH's are due to just one 
type of isometry: the Clifford translations (CT)~\cite{GTRB1998}, which 
are isometries $g_t \in \Gamma $ such that for all $p \in \widetilde{M}$ 
the distance $d(p, g_t p)$ is a constant (see also in this regard%
~\cite{LeLuUz1999}). Clearly the CT's reduce to the regular translations 
in the Euclidean spaces (for more details and simulations see~\cite{GRT2000a}~--~\cite{GRT2001a}).
Since there is no CT translation in hyperbolic geometry this 
result explains the absence of spikes in the PSH's of hyperbolic 
universes with nontrivial detectable topology. On the other hand, 
it also makes clear that distinct manifolds which admit the same 
Clifford translations in their covering groups present the same spike
spectrum of topological origin. Therefore the topological spikes 
are not sufficient for unambiguously determine the topology of the 
universe.

In spite of these limitations, the most striking evidence of 
multiply-connectedness in PSH's is indeed the presence of topological 
spikes, which result from translational isometries $g_t \in \Gamma\,$.
It was demonstrated~\cite{GTRB1998,GRT2000a} that the other isometries $g$ 
manifest as very tiny deformations of the expected pair separation 
histogram $\Phi^{sc}_{exp}(s_i)$ corresponding to the underlying 
simply connected universe~\cite{BernuiTeixeira1999,Reboucas2000}.
Furthermore, in PSH's of universes with nontrivial topology the 
amplitude of the sign of non-translational isometries was shown to be
smaller than the statistical noise~\cite{GRT2000a}, making clear
that one cannot use PSH to reveal these isometries.

In brief, the only significant (measurable) sign of a nontrivial 
topology in PSH are the spikes, but they can be used merely to 
disclose (not to determine) a possible nontrivial topology 
of universes that admit Clifford translations: any flat, some 
spherical, and no hyperbolic universes. 

The impossibility of using the PSH method for the detection of 
the topology of hyperbolic universes motivated the development 
of a new scheme called \emph{collecting correlated pairs method} 
(CCP method)~\cite{UzLeLu1999} to search for cosmic topology. 

In the CCP method it is used the basic feature of the isometries, 
i.e., that they preserve the distances between pairs of images.
Thus, if $(p,q)$ is a pair of arbitrary images (correlated or not)
in a given catalogue $\mathcal{C}$, then for each $g \in \Gamma$ 
such that the pair $(gp,gq)$ is also in $\mathcal{C}$ we obviously 
have
\begin{equation}   \label{typeI}
d(p,q) = d(gp,gq) \; .
\end{equation}
This means that for a given (arbitrary) pair $(p,q)$ of images 
in $\mathcal{C}$, if there are $n$ isometries $g \in \Gamma$ such 
that both images $gp$ and $gq$ are still in $\mathcal{C}$, then
the separation $s(p,q)$ will occur $n$ times. 

The easiest way to understand the CCP method is by looking into  
its computer-aimed procedure steps, and then examine the consequences 
of having a multiply connected universe with detectable topology. 
To this end, let $\mathcal{C}$ be a catalogue with $n$ sources, 
so that one has $P = n(n-1)/2$ pairs of sources. The CCP procedure 
consists on the following steps:
\begin{enumerate}
\item 
Compute the $P$ separations $s(p,q)$, where $p$ and $q$ are two images 
in the catalogue $\mathcal{C}$;
\item 
Order the $P$ separations in a list $\{s_i\}_{1 \leq i \leq P}$ such 
that $s_i \leq s_{i+1} \,$; 
\item 
Create a list of \emph{increments} $\{ \Delta_i \}_{1 \leq i \leq P-1}$, 
where $\Delta_i = s_{i+1} - s_i \,$;.
\item 
Def\/ine the CCP index as
\begin{displaymath}
\mathcal{R} = \frac{\mathcal{N}}{P-1} \, ,
\end{displaymath}
where $\mathcal{N} = Card\{i \,:\, \Delta_i = 0\}$ is the
number of times the increment is null.
\end{enumerate}

If the smallest characteristic length of $M$ exceeds the survey 
depth ($r_{inj} > \chi_{obs}$) the probability that two pairs of 
images are separated by the same distance is zero, so 
$\mathcal{R} \approx 0$. On the other hand, in a universe with
detectable nontrivial topology ($\chi_{obs}> r_{inj}$) given 
$g \in \Gamma$, if $p$ and $q$ as well as $gp$ and $gq$ are 
images in $\mathcal{C}$, then: 
(i) the pairs $(p,q)$ and $(gp,gq)$ are separated by the same 
distance; and 
(ii) when $\Gamma$ admits a translation $g_t$ the pairs $(p,g_t p)$ 
and $(q,g_t q)$ are also separated by the same distance.
It follows that when a nontrivial topology is detectable, and
a given catalogue $\mathcal{C}$ contains multiple images, then 
$\mathcal{R} > 0$, so the CCP index is an indicator of a 
detectable nontrivial topology of the spatial section $M$ of 
the universe. Note that although $\mathcal{R}> 0$ can be used 
as a sign of multiply connectedness, it gives no indication as 
to what the actual topology of $M$ is. Clearly whether one 
can find out that $M$ is multiply connected (compact in at 
least one direction) is undoubtedly a very important step, 
though. 

In more realistic situations, uncertainties in the determination
of positions and separations of images of cosmic sources are 
dealt with through the following extension of the CCP index:%
~\cite{UzLeLu1999} 
\begin{displaymath}
\mathcal{R_{\epsilon}} = \frac{\mathcal{N_{\epsilon}}}{P-1} \; ,
\end{displaymath}
where $\mathcal{N_{\epsilon}} = Card\{i \, : \, \Delta_i \leq
\epsilon\}$, and $\epsilon > 0$ is a parameter that quantifies
the uncertainties in the determination of the pairs separations.

Both PSH and CCP statistical methods rely on the accurate 
knowledge of the three-dimensional positions of the cosmic 
sources. The determination of these positions, however, 
involves inevitable uncertainties, which basically arises 
from:
(i) uncertainties in the determination of the values of the
cosmological density parameters $\Omega_{m0}$ and
$\Omega_{\Lambda 0}$; 
(ii) uncertainties in the determination
of both the red-shifts (due to spectroscopic limitations),
and the angular positions of cosmic objects (displacement,
due to gravitational lensing by large scale objects, e.g.); and 
(iii) uncertainties due to the peculiar velocities of 
cosmic sources, which introduce peculiar red-shift 
corrections.
Furthermore, in most studies related to these methods the catalogues 
are taken to be complete, but real catalogues are incomplete: objects 
are missing due to selection rules, and also most surveys are not 
full sky coverage surveys.
Another very important point to be considered regarding these
statistical methods is that most of cosmic objects do not have 
very long lifetimes, so there may not even exist images of a
given source at large red-shift. This poses the important problem 
of what is the suitable source (candle) to be used in these 
methods. 

Some of the above uncertainties, problems and limits of the 
statistical methods have been discussed by Lehoucq \emph{et al.\/}%
~\cite{LeUzLu2000}, but the robustness of these methods still 
deserves further investigation. So, for example, a quantitative 
study of the sensitivity of spikes and CCP index with respect to 
the uncertainties in the positions of the cosmic sources, which 
arise from unavoidable uncertainties in values of the density 
parameters is being carried out~\cite{BeGoMoRe2004}. 

For completeness we mention the recent articles by Marecki \emph{et al.}%
~\cite{MareckiRoukemaBajtlik}, and by Bernui \emph{et al.}~\cite{Bernui2003}.
Bernui and Villela have worked with a method which uses pair angular 
separation histograms (PASH) in connection with both discrete cosmic sources 
and CMBR.

To close this section we refer the reader to references%
~\cite{RoukemaEdge1997,FagGaus1999}, which present alternative  
statistical methods (see also the review articles~\cite{Reviews}).

\section{Looking for the Topology using CMBR}
\label{CMBmethods}

The CMB temperature anisotropy measurements by WMAP combine high 
angular resolution, and high sensitivity, with the full sky and the deepest
survey ($z_{LSS} \sim 1\,100$) currently available. These features make 
very promising the observational probe of cosmic topology with CMBR 
anisotropies on length scales near to or even somewhat beyond the 
horizon $\chi_{hor}$.

Over the past few years distinct approaches to probe a non-trivial 
topology of the universe using CMBR have been suggested. In a recent
paper Souradeep and Hajian~\cite{HajianSouradeep05,SouradeepHajian05}  
have grouped these approaches in three broad families. Here, however, we
shall briefly focus on the most well known method that relies on multiple images 
of spots in the CMBR maps, which is known as circles-in-the-sky%
~\cite{CSS1998} (for more detail on the other methods see, e.g.,
refs.~\cite{Levin98}~--~\cite{Dopi04}).

For an observer in the Hubble flow the last scattering surface (LSS) 
is well approximated by a two-sphere of radius $\chi_{LSS}$. If a 
non-trivial topology of space is detectable, then this sphere intersects 
some of its topological images, giving rise to circles-in-the-sky, i.e., 
pairs of matching circles of equal radii, centered at different points 
on the LSS sphere, with the same pattern of temperature variations%
~\cite{CSS1998}.
These matched circles will exist in CMBR anisotropy maps of universes 
with any detectable nontrivial topology, regardless of its geometry. 

The mapping from the last scattering surface to the sky sphere is
conformal. Since conformal maps preserve angles, the identified 
circle at the LSS would appear as identified circles on the sky sphere.
A pair of matched circles is described as a point in a six-dimensional 
parameter space. These parameters are the centers of each circle, which are 
two points on the unit sphere (four parameters), the angular radius of both 
circles (one parameter), and the relative phase between them (one parameter). 

Pairs of matched circles may be hidden in the CMBR maps if the universe has
a detectable topology. Therefore to observationally probe nontrivial topology
on the available largest scale, one needs a statistical approach to scan all-sky 
CMBR maps in order to draw the correlated circles out of them.%
\footnote{In practice, however, contributions such as integrated Sachs-Wolfe
effect, the tickness of the last scattering surface, and even aberration effect%
~\cite{Calvaoetal} may damage or even destroy the matching of the
circles.}
To this end, let $\mathbf{n}_1=(\theta_1,\varphi_1)$ and $\mathbf{n}_2=(\theta_2, 
\varphi_2)$ be the center of two circles $C_1$ and $C_2$ with angular radius 
$\rho$. The search for the matching circles can be performed by computing the 
following correlation function~\cite{CSS1998}:
\begin{equation}
\label{CirSky}
S(\alpha) = \frac{\langle 2 T_1(\pm \phi) T_2(\phi + \alpha) 
\rangle}{\langle T_1(\pm \phi)^2 + T_2(\phi + \alpha)^2 \rangle} 
\; ,
\end{equation}
where $T_1$ and $T_2$ are the temperature anisotropies along 
each circle, $\alpha$ is the relative phase between the two circles, 
and the mean is taken over the circle parameter $\phi\,$: 
$\langle \;\, \rangle = \int_{0}^{2\pi} d\phi$. The plus $(+)$ and
minus $(-)$ signs in (\ref{CirSky}) correspond to circles correlated, 
respectively, by non-orientable and orientable isometries.
 
For a pair of circles correlated by an isometry (perfectly matched) 
one has $T_1(\pm \phi) = T_2(\phi + \alpha_*)$ for some 
$\alpha_*$, which gives $S(\alpha_*) = 1$, otherwise the circles 
are uncorrelated and so $S(\alpha) \approx 0$. 
Thus a peaked correlation function around some $\alpha_*$ would 
mean that two matched circles, with centers at $\mathbf{n}_1$ and $
\mathbf{n}_2$, and angular radius $\rho$, have been detected. 

{}From the above discussion it is clear that a full search for 
matched circles requires the computation of $S(\alpha)$, for any 
permitted $\alpha$,  sweeping the parameter sub-space 
$(\theta_1 , \varphi_1, \theta_2 , \varphi_2, \rho)$, and so it 
is indeed computationally very expensive. Nevertheless, such 
a search is currently in progress, and preliminary results 
using the first year WMAP data failed to find antipodal and 
nearly antipodal, matched circles with radii larger than $25^\circ$%
~\cite{CSSK2003}. Here nearly antipodal means circles whose centers are 
separated by more than $170^\circ$. At a first sight this preliminary 
result seems to rule out topologies whose isometries produce antipodal 
images of the observer, as for example the Poincar\'e dodecahedron 
model~\cite{Poincare}, or any other homogeneous spherical space with 
detectable isometries. In this regard, it is important to note the 
results of the recent articles by Roukema \emph{et al.}~\cite{Roukema} 
and Aurich \emph{et al.}~\cite{Aurich}, Gundermann~\cite{Gundermann}, 
and some remarks by Luminet~\cite{Luminet05}, which support the 
dodecahedron model.

Furthermore, since detectable topologies (isometries) do not produce,
in general, antipodal correlated circles, a little more can be inferred from
the lack of nearly antipodal matched circles. Thus, in a flat universe,
e.g., any screw motion may generate pairs of circles that are not even 
nearly antipodal, provided that the observer's position is far enough 
from the axis of rotation~\cite{Gomero2003b,RiazUzLeLu2003}. 
As a consequence, our universe can still have a flat topology, other than 
the $3$-torus, but in this case the axis of rotation of the screw motion 
corresponding  to a pair of matched circles would pass far from our position,
making clear the crucial importance of the position of the observer relative 
to the 'axis of rotation' in the matching circles search method.

\begin{theacknowledgments}
I thank CNPq for the grant under which this work was carried out,
and M. Novello for the invitation to give a set of lectures in the
$\mathrm{XI}^{\mathrm{th}}$ Brazilian School of Cosmology and Gravitation. 
I also thank B.\ Mota for his valuable help in polishing the figures
and for the reading of the manuscript and indication of misprints 
and omissions. 

In my life, learning has often been a cooperative process in which contributions 
have come from many quarters consciously as well as unconsciously. I have 
learned a lot from my collaborators. Many warmest thanks go to them 
all.
\end{theacknowledgments}

\end{document}